\documentclass[a4paper,twocolumn,english,aps,prl]{revtex4}

\usepackage{amsfonts}
\usepackage{amsmath}
\usepackage{amssymb}
\usepackage{graphicx,color}
\usepackage{amsmath}
\providecommand{\U}[1]{\protect\rule{.1in}{.1in}}
\begin{document}
\preprint{cond/mat}
\title[FCS as Geometry]{Full Counting Statistics as the Geometry of Two Planes}
\author{Y.B. Sherkunov, A. Pratap,  B. Muzykantskii, N. d'Ambrumenil}
\affiliation{Department of Physics, University of Warwick, Coventry CV4 7AL, United Kingdom}
\pacs{PACS number}

\begin{abstract}
Provided the measuring time
is short enough, the full counting statistics (FCS) of 
the charge pumped across a barrier as a result of 
a series of voltage pulses 
are shown to be equivalent to the geometry of two planes.
This formulation leads to the FCS
without the need for the usual non-equilibrium (Keldysh) transport
theory or the direct computation of the determinant of an infinite-dimensional
matrix. In the particular case of the application of $N$ 
Lorentzian pulses, we show the computation of the FCS reduces to
the diagonalization of an $N\times N$ matrix.
We also use the formulation to compute the core-hole response in the 
X-ray edge problem and the FCS for a square wave pulse-train for the case 
of low transmission.

\end{abstract}
\volumeyear{year}
\volumenumber{number}
\issuenumber{number}
\eid{identifier}
\date{\today}
%\received[Received text]{date}

%\revised[Revised text]{date}

%\accepted[Accepted text]{date}

%\published[Published text]{date}

\startpage{1}
\endpage{ }
\maketitle
%\tableofcontents
The  experimental and technological importance of 
a full quantum mechanical treatment of the  response of 
Fermi systems to a time-dependent perturbation, has grown
as electronic devices have shrunk. In particular, the quantum statistics
of charge transfer events induced by an applied voltage
pulse in simple systems such as across a tunneling
barrier or along a wire need to be understood if the
limits to their capacity to encode or transmit information are to
be found \cite{Beenakker2001,LevitovReznikov04}. 
They are also excellent examples of non-equilibrium systems for which
in some cases complete theoretical treatments are known or are
feasible. 

Theoretically, interest has concentrated on 
the full counting statistics (FCS) or generating function,  $\chi(\lambda)$, 
for all moments of the 
charge distribution \cite{LevitovLeeLesovik96}.
There are
results for the case of an infinite train of periodic pulses impinging on a 
tunneling barrier \cite{IvaLL97,dAMuz05,VanevicNazBelz07} and, 
for the case of a low transparency barrier
with a constant bias voltage $V$, the FCS are also known 
for non-zero temperatures \cite{LevitovReznikov04}.
%FCS of the more general
%case of a combination of Lorentzian pulses of both polarities 
%has also been analyzed at zero temperature and 
%expressions for the shot noise contribution to the 
Particular attention has been paid to the case
of Lorentzian voltage pulses with `integer area'
($\frac{e}{h}\int_{-\infty}^\infty V(t) dt$  is an integer).
If the voltage pulses all have the same polarity they generate
so-called `Minimal excitation states' (MES). These have been
shown to minimise the number of
excitations in a one-channel Fermi gas required to generate a given signal.
They  have the intriguing property that a train of such pulses retain
the minimal noise property \cite{IvaLL97,KeelingKlichLev06}
independent of the separation or width of the generating voltage pulses
provided the pulses are all of the same polarity.
When such signals impinge on a tunneling barrier  
their nature is reflected in the FCS \cite{IvaLL97}. 

\begin{figure}[t]
%\includegraphics[width=3.5in,
%  height=4.7in ]{bup.eps}
%\input{rotated.pstex_t}
\includegraphics{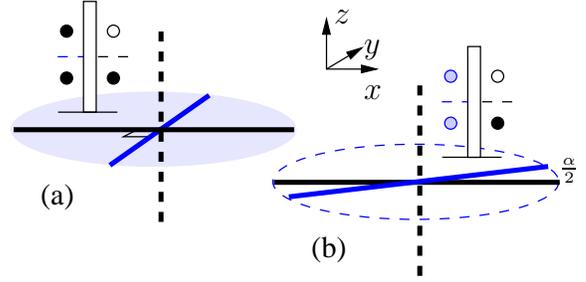}
\caption{\label{fig:rotation} (color online)
The geometry of the FCS.
In the space of single-particle states, orbitals occupied at zero temperature
correspond to eigenstates of $h$ with eigenvalue $1$ and
define a  mirror plane 
shown schematically as the $x-$axis. 
Unoccupied states at zero temperature (with eigenvalue $-1$) are in the complement
of plane, one of these is the y-axis and the remainder are shown as the z-axis. In the
insets the dashed line divides states into those above and below the unperturbed
Fermi level but no other ordering by energy is implied.
Application of a voltage pulse, biasing one electrode with respect to
the other, transforms $h \rightarrow \widetilde{h} = UhU^\dagger$.
Case (a): ($N_u=1$, $N_b=0$ see text) The corresponding plane has an added dimension
and is shown as the shaded plane.
Case (b):  ($N_u=0$, $N_b=1$) The plane rotates
by $\alpha/2$ about an axis perpendicular to $h$. The axis of rotation and the
new occupied state are found from $[h,e^{-i\phi(t)}]$ (see text). All other states
are rotated as a function of time but remain eigenstates of {\bf both} $h$ 
and $\widetilde{h}$. }  
\end{figure}

Here we show that the FCS at zero temperature
are determined solely by the geometry of two planes.
For a train of $N$ Lorentzian pulses these two planes
coincide except in an $N$-dimensional subspace. Computing the FCS in this case 
reduces to the diagonalization of $N\times N$ matrix, which
we compute explicitly.
The known results for $\chi(\lambda)$ follow directly
from this geometric formulation circumventing the need
to work with a formulation using
Keldysh Green's functions \cite{VanevicNazBelz07} 
or solving an auxiliary Riemann-Hilbert problem to invert singular 
operators \cite{dAMuz05}.

The mirror planes are defined by
the operators $h=2n-1$ and $\tilde{h} = UhU^\dagger$, where
$n$ is the ground state density matrix and $U$ is
the unitary operator which describes the effect of applying a voltage pulse
between different components of the sytem. Directions corresponding to
occupied single-particle states are in the corresponding mirror plane (eigenvalue $+1$) while
directions corresponding to unoccupied states are in its complement and reflected
(eigenvalue $-1$).
The planes have the dimension of the number 
of occupied single-particle states.
The intersection
of the two planes (and the intersection of the two complements)
define the states which are occupied (or unoccupied) in {\bf both} the initial and transformed states
and which therefore contribute nothing to the FCS. The 
remaining states are rotated
with respect to one another by angles which we denote by $\frac{\alpha_k}{2}$. 
The directions which are rotated can be 
identified from the commutator $[h,e^{-i\phi(t)}]$, which defines a spanning set of states for the
rotated space.

We consider as a model system a quantum point contact between two ideal single
channel conductors coupled by a tunneling barrier localised around $x=0$
and treat the effect of a voltage pulse, $V(t)$, applied between
the two conductors (see Figure \ref{fig:rotation}). We assume that we can neglect the 
contributions to
the noise and other moments of the distribution of transferred
electrons which 
scale as $\log{t_0 \xi }$ where $t_0$ is the observation time and $\xi$
is an ultra-violet cutoff set by the Fermi energy. These are present 
even in equilibrium and are associated with the fact that the charge number in the left or
right electrode is not a conserved quantity \cite{LevitovLeeLesovik96}. 
The FCS
of transmitted charge across the point contact (which we take to have reflection and 
transmission probabilities $R$ and $T=1-R$) can be
obtained from the generating function 
\begin{equation}
\chi(\lambda) = \sum_{n=-\infty}^{\infty} P_n e^{i\lambda n} 
\label{eq:GeneratingFunction}
\end{equation}
where $P_n$ is the probability of $n$ particles being transmitted
(from the left electrode to the right one).

The principle result of \cite{VanevicNazBelz07} is that 
\begin{eqnarray} 
\chi(\lambda) & = & (R + Te^{i\kappa_l\lambda})^{N_u} 
\nonumber \\
& & \prod_{1}^{N_b} 
\left(1 + R T\sin^2{\frac{\alpha}{2}}(e^{ i \lambda} + e^{- i \lambda}-2)\right),
\label{eq:VanNazBel}
\end{eqnarray}
where $N_u$ is the number of {\it unidirectional\/} events with
$\kappa_l=\pm 1$ determined by the polarity of the voltage pulse and 
$N_b$ the number of {\it bidirectional \/} events. 
The geometry for the two types of event is illustrated in Figure \ref{fig:rotation}.
Consider first the case of a simplified system consisting of a single state
on either side of the barrier. 
Suppose now that, after
taking account of the voltage pulse,  the state on the left side is occupied
and empty on the right. This is an example of a single MES \cite{KeelingKlichLev06}
or {\it unidirectional\/} \cite{VanevicNazBelz07} event and is
the case illustrated in 
Fig \ref{fig:rotation}(a): the $h$-plane (shown as the $x-$axis) is actually
zero-dimensional for this
simplified one-state system and the $\widetilde{h}$ plane is one-dimensional.
The FCS for this situation is clearly given by $\chi(\lambda) = R + Te^{i\lambda}$:
$P_0=R$ and $P_1=T$ in (\ref{eq:GeneratingFunction}).

The {\it bidirectional \/} event is illustrated in Fig \ref{fig:rotation}(b). 
Again consider first the case
of a simplified system of a single occupied state on either side of the barrier. If
the occupied states are equivalent ({\it i.e.\/}
they have the same energy or the same decomposition by energy),
then transfer across the barrier is blocked and $\chi(\lambda) = 1$. If the two
states are orthogonal to one another (have different energies or orthogonal decompositions
by energy) then each state behaves independently. The FCS for this case
is a product over that for two independent events
each of which is of the type illustrated in Fig \ref{fig:rotation}(a). This would give
$\chi(\lambda) = (R + Te^{i\lambda})(R + Te^{-i\lambda})$. In general, 
the projection of the two occupied states onto one another is 
equal to  $\cos^2{\frac{\alpha}{2}}$, where
$\frac{\alpha}{2}$ is the angle between the two planes in Fig \ref{fig:rotation}. 
Now, the FCS for this simplified system is just the weighted average
of the two previous cases:
$\cos^2{\frac{\alpha}{2}} + \sin^2{\frac{\alpha}{2}}(R + Te^{i\lambda})(R + Te^{-i\lambda})
= 1 + RT\sin^2{\frac{\alpha}{2}} (e^{i\lambda} + e^{-i\lambda} - 2)$.

The general case of a filled Fermi sea is a combination of the two types of events
illustrated in Figure \ref{fig:rotation}. This follows from the fact
that the operator $h\widetilde{h}$ is unitary and therefore has an orthonormal
basis which can be contructed from its eigenvectors \cite{VanevicNazBelz07}.
For positive mean transfer from left to right electrode $N_u>0$ \cite{note_on_integer_Nu},
the dimensionality of $\widetilde{h}$ is $N_u$ greater than that of $h$. There 
are therefore $N_u$ directions with eigenvalue of $h\widetilde{h}$ equal to -1. 
The remaining eigenvalues come in pairs ($e^{\pm i\alpha_j}$ with $j=1\dots N_b$)
and the 2D plane defined by the corresponding basis vectors, $e_{2j-1} \times e_{2j}$,  
intersects the two mirror-planes defined by $h$ and $\widetilde{h}$ in two
lines inclined at an angle $\alpha_j/2$.  The FCS, $\chi(\lambda)$, is then
given by the product over the appropriate factors for the $N_u$ and $N_b$ events
to give (\ref{eq:VanNazBel}).

The case where $N_b$ and $N_u$ are both finite corresponds to a train of
Lorentzian pulses \cite{note_on_integer_Nu} and has 
attracted a lot of attention \cite{IvaLL97,KeelingKlichLev06}. Here we show
that it is possible to identify explicitly the
sub-space for which the eigenstates of $h \widetilde{h}$ have eigenvalues different
from unity. 
%The case of $N_b=1$, $N_u=0$ is by far the easiest non-trivial case
%to handle and the FCS can be computed analytically.
%It is made simple by the fact only two states are involved and these are automatically
%orthogonal. In the general case t
The computation of the FCS reduces to the 
diagonalization
of an $N \times N$ matrix where $N = N_u + 2N_b$ to find the angles $\alpha_i$ from
its eigenvalues ($e^{i\alpha_i}$).
The flux through the circuit, $\phi(t) =  -(e/h)\int^t V(t') dt'$, 
can be written in this case as
\begin{equation}
e^{i\phi(t)}  =  \prod_{m=1}^N \frac{t- p_m^*}{t-p_m} = 1 + \sum_{m=1}^N \frac{A_m}{t-p_m}.
\label{eq:quantizedpulses} 
\end{equation}
The poles at $p_m = t_m + i \tau_m$ have residues $A_m$, which are determined by
the system of equations 
\begin{equation}
-1  =  \sum_{m=1}^N \frac{A_m}{p^*_n-p_m},  %=  \sum_{m=1}^N \frac{A^*_m}{p_n-p_m^*}, 
\hspace{0.5cm} n=1,\ldots, N.
\label{eq:zeros}
\end{equation}
%We will assume that $N=N_u+2N_b$ with $\tau_m<0$ for $m=1,\ldots,N_b$,  and $\tau_m>0$
%for $m=(N_b+1),\ldots,N$.
To within overall phase factors, the effect of the voltage pulse is
equivalent to a multiplication of left electrode states by the factor $e^{i\phi(t)}$. 

The states which define the space in which the eigenvalues of $h\widetilde{h}$ are different from 1
are $|\psi_m(t) \rangle = 1/(t-p_m)$ for $m=1,\ldots,N$. This follows from considering 
explicitly the effect of $U$.
All single particle states, $\psi$, satisfying $(h \widetilde{h}-1)|\psi\rangle  =
 he^{i\phi}[h,e^{-i\phi}]| \psi \rangle = 0$, do
not change occupancy under the effect of $U$ (and hence make no contribution to
$\chi(\lambda)$).
For an arbitrary single-particle state $\psi(t)$, we find
using (\ref{eq:quantizedpulses}) and (\ref{eq:zeros})
\begin{equation}
(h \widetilde{h} - 1) |\psi\rangle =   \sum_{m,m'}
 \mbox{sign}(\tau_{m'})\frac{  2\, A_{m'}A^*_m }{p^*_{m}-p_{m'} }
\frac{f_m}{t-p_{m'}}.
\label{eq:commutator}
\end{equation} 
Here $f_{m}$ is the projection of $\psi(t)$ onto $|\psi_m(t)\rangle$:
\begin{equation}
f_{m} = \frac{i}{2\pi}\int dt \frac{1}{(t-p_m)^*} \psi(t).
\label{eq:project_on_pm*}
\end{equation} 
The right-hand side of (\ref{eq:commutator}) vanishes for all 
$\psi(t)$ orthogonal to the set of states $\{\psi_n(t), n=1, \ldots N\}$. 

The FCS for the case of a train of $N$ Lorentzian pulses
can then be found by considering the effect of $h\widetilde{h}$ on the 
subspace of the states  $\{\psi_n(t)\}$. We find
$h\widetilde{h} |\psi_n\rangle = \sum_{m}M_{m n}|\psi_m\rangle$, where:
\begin{equation}
M_{mn} =  \sum_{m'} 
\frac{\mbox{sign}(\tau_m \tau_{m'}) A_m A_{m'}^*}{(p_m-p_{m'}^*)(p_n-p_{m'}^*)}.
\label{eq:Mmn}
\end{equation}
The angles $\alpha_i$ which determine the FCS in (\ref{eq:VanNazBel}) are found
from the eigenvalues ($e^{i\alpha_i}$) of the $N\times N$ matrix $M_{mn}$.

Analytic results for the FCS are known in some simple cases
\cite{IvaLL97}: i) The $\{\tau_n\}$ all have the same sign and 
ii) $N=2$ with $\mbox{sign}(\tau_1\tau_2) = -1$. The case
of $N=4$ has also been solved analytically \cite{Pratap08}.
Case i) corresponds to the case of MES. All poles of $e^{i\phi}$ are
in the same half-plane and all $N$ states, $|\psi_n(t)\rangle = 1/(t-p_m)$
are eigenstates of $h\widetilde{h}$ with eigenvalue -1. This can be seen
from (\ref{eq:Mmn}) using  $\mbox{sign}(\tau_m \tau_m')=1$,  (\ref{eq:zeros})
and by considering $\frac{de^{-i\phi(t)}}{dt}$.
These states define the $N$ non-coplanar directions which are
in one mirror plane and in the complement of the other.
The FCS for this case of MES corresponds to $N_u = N$ {\it unidirectional\/} events in
(\ref{eq:VanNazBel}).

Case ii) corresponds to a single {\it bidirectional\/} event. Here the FCS are just
found from the eigenvalues of the $2\times 2$ matrix $M_{mn}$.
Explicit calculation gives the result $\sin^2\alpha/2 = | \frac{p_1 - p_2^*}{p_1-p_2} |^2$
\cite{IvaLL97}. In this case, because the states  $|\psi_1\rangle$ and $|\psi_2\rangle$
are orthogonal to each other, it is also particularly simple to compute the effect
of $U$ on the single-particle states explicitly. 
We work with normalized states $|\bar{\psi}_i\rangle = C_i |\psi_i\rangle$ where
$C_i= \sqrt{|p_i-p_i^*|}$. We find that the
effect of $U$  on the state $|\bar{\psi}_1^* \rangle = C_1/t-p_1^*$ is to
transform it into a linear combination of $\bar{\psi}_1$ and $\bar{\psi}_2$:
\begin{equation}
    e^{i\phi(t)} |\bar{\psi}_1^*\rangle  =  
\frac{C_1(p_2^*- p_2)}{C_2(p_1-p_2)} |\bar{\psi}_2\rangle
+ \frac{p_1 - p_2^*}{p_1-p_2} |\bar{\psi}_1\rangle. 
\label{eq:1in1out}
\end{equation}
If $|\psi_1^*\rangle$ is in the mirror plane corresponding to $h$ then 
the component of the transformed state proportional to 
$|\psi_1\rangle$ is in the complement of the mirror plane and its 
amplitude is given by $\sin {\frac{\alpha}{2}} e^{i\phi_1} =  \frac{p_1 - p_2^*}{p_1-p_2}$
a result previously obtained using an operator formalism \cite{KeelingKlichLev06}.

In the general case
when $N_b$ is not finite \cite{note_on_integer_Nu}, the geometric
approach simplifies calculations when the transparency, $T$, 
(or the scattering phase shifts in
other problems)
are low, it is possible to deduce the results without the need for
any direct diagonalization. 
We illustrate this by rederiving the standard result for the orthogonality
catastrophe or Fermi edge singularity (FES) problem and then use the method
to generalise results for the FCS for a system subjected to a sudden pulse of finite duration.

The FES problem arises in the context of the X-ray absorption
spectrum of a metal. Here, the form of the spectrum is determined by the 
response of the conduction electrons to the sudden appearance of a core hole
\cite{ND69} and has been shown to be the consequence of the orthogonality of the
ground states of the Fermi gas with and without the local potential due to the
core hole \cite{Anderson67}. The core hole spectral function is the Fourier transform  with
respect to $t_f$ of the overlap between the state, $U(t_f) |0\rangle $, where
$|0\rangle$ is 
initial ground state
at $t=0$ and $U(t_f)$ the unitary time-development operator of the system. $U(t_f)$
desribes the effect of the new potential which switches on at $t=0$.
(All states are written in the interaction basis
with $H_0$ the Hamiltonian of the unperturbed system.) This is a one-channel version
of the problem we have considered. The overlap 
$\langle 0 | U | 0 \rangle = \prod_k \cos{\alpha_k/2}$, where the $e^{\pm i \alpha_k}$
are the eigenvalues of $h \widetilde{h}$. 
When the rotation angles $\alpha_k$ are 
small, we can expand the cosine to obtain 
$\langle 0 | U | 0 \rangle \approx 
\exp{ \left( \frac{1}{8} \mbox{Tr} (h \widetilde{h} - 1 )  \right)  }$.
In the usual one-channel FES problem, $U = e^{i\phi(t)}$ where 
$\phi(t)=2\phi_0 \theta(t) \theta(t_f-t)$ 
and where $\phi_0$ is the phase shift of the core hole potential computed on the scattering (unperturbed)
states of the system. If we work in
the time-domain, $h(t,t')=\frac{i}{\pi} \mbox{P}(\frac{1}{t-t'})$ where $P$ denotes
Cauchy principal part,  we find
\begin{eqnarray}
\mbox{Tr}\left(h \widetilde{h} - 1 \right) & = & \int dt dt' h(t,t') 
\left(e^{i\phi(t)}e^{-i\phi(t')} - 1\right) h(t',t) \nonumber \\
& = & -\frac{8}{\pi^2} \sin^2{\phi_0} \log{\xi t_f},
\label{eq:overlap}
\end{eqnarray}
($\xi$ is the usual  short-time cutoff of order the Fermi energy) and
we recover the 
standard result for this problem: $\langle 0 | U | 0 \rangle = 
(\xi t_f)^{-\phi_0^2/\pi^2}$.

The FCS for a train of step pulses in the low transparency limit is known to be
Poissonian \cite{LevitovJETP93}. Here we will consider a periodic signal with pulses of 
length $t_p$ and
period $S$ applied to the left electrode with respect
to the right electrode: $V(t)=2 \phi_0 (\delta(t-nS) - \delta(t-nS-t_p))$. 
The geometry of this situation is very similar
to that of the  core hole Green's function in the FES problem. 
The general result (\ref{eq:VanNazBel}) then
gives
$\log \chi(\lambda) = \sum_k \log[1+TR\sin^2 \alpha_k/2 (e^{i\lambda} + e^{-i\lambda} - 2)]$. In the
limit of $TR\ll 1$, we can expand the logarithm and use
$\sum_k \sin^2 \frac{\alpha_k}{2} = - \frac{1}{4} \mbox{Tr} (h \widetilde{h} - 1) $. We find
\begin{eqnarray}
\mbox{Tr} \left(h\widetilde{h}-1\right) & = & - \frac{4}{\pi^2} \sum_{m,n} 
\left[\log{ \left( \frac{t_p+S(m-n-1)}{S(m-n-1)+\tau}  \right) } \right. \nonumber \\
& & \left. + \log{ \left( \frac{t_p+S(m-n)}{S(m-n)+\tau}  \right) } \right] \sin^2{\phi_0} ,
\label{eq:trace_for_step_pulse}
\end{eqnarray}
where $\tau$ is short time cut-off used to characterise the delta-function
$\delta(t) = \frac{\tau}{\pi(t^2+\tau^2)}$ and 
$N_u=0$ for this periodic train of pulses. In (\ref{eq:trace_for_step_pulse}), 
the terms with $m=n$ and $m=n+1$  are divergent for $\tau \rightarrow 0$, 
while all other terms are convergent.
With a measurement time $N S$ we obtain the usual result 
$\log \chi(\lambda) = 2N \frac{RT}{\pi^2} 
\sin^2\phi_0 \log\frac{t_p}{\tau} (e^{i \lambda} + e^{-i \lambda} -2)$.

Finally we comment on the logarithmic contributions to the noise which are present
even in the absence of an applied voltage pulse and limit possible applications and the 
observability of minimal noise states. 
When calculating the FCS, the system is assumed to have been prepared
in the ground state of the system with zero transmission $T=0$. 
The scattering matrix is taken to be different from unity only during
the observation time $t_0$ when the transmission becomes non-zero ($T>0$) and
the number of particles on a particular
side of the barrier is no longer a good quantum number. The geometry of
this situation is similar to the case of the step pulse in voltage except 
that the unitary transformation acting on the initial states mixes states on both
sides of the barrier \cite{dAMuz08}. In practice, these logarithmic corrections
imply that the observation time $t_0$ for the MES and other effects discussed here to be observable
above the equilibrium noise must be short.
For a system with $\epsilon_F = 10$meV and setting the equilibrium noise 
at low temperatures
($\frac{1}{\pi^2}\log{\epsilon_F t_0/h}$) equal to the minimal noise obtained using from a MES
corresponds to working at frequencies $\nu > 300$ MHz.

\bibliographystyle{apsrev}
\bibliography{thermalnoise}

\begin{thebibliography}{13}
\expandafter\ifx\csname natexlab\endcsname\relax\def\natexlab#1{#1}\fi
\expandafter\ifx\csname bibnamefont\endcsname\relax
  \def\bibnamefont#1{#1}\fi
\expandafter\ifx\csname bibfnamefont\endcsname\relax
  \def\bibfnamefont#1{#1}\fi
\expandafter\ifx\csname citenamefont\endcsname\relax
  \def\citenamefont#1{#1}\fi
\expandafter\ifx\csname url\endcsname\relax
  \def\url#1{\texttt{#1}}\fi
\expandafter\ifx\csname urlprefix\endcsname\relax\def\urlprefix{URL }\fi
\providecommand{\bibinfo}[2]{#2}
\providecommand{\eprint}[2][]{\url{#2}}

\bibitem[{\citenamefont{Beenakker and Schomerus}(2001)}]{Beenakker2001}
\bibinfo{author}{\bibfnamefont{C.~W.~J.} \bibnamefont{Beenakker}}
  \bibnamefont{and}
  \bibinfo{author}{\bibfnamefont{H.}~\bibnamefont{Schomerus}},
  \bibinfo{journal}{Phys. Rev. Lett.} \textbf{\bibinfo{volume}{86}},
  \bibinfo{pages}{700} (\bibinfo{year}{2001}).

\bibitem[{\citenamefont{Levitov and Reznikov}(2004)}]{LevitovReznikov04}
\bibinfo{author}{\bibfnamefont{L.~S.} \bibnamefont{Levitov}} \bibnamefont{and}
  \bibinfo{author}{\bibfnamefont{M.}~\bibnamefont{Reznikov}},
  \bibinfo{journal}{Phys. Rev. B} \textbf{\bibinfo{volume}{70}},
  \bibinfo{pages}{115305} (\bibinfo{year}{2004}).

\bibitem[{\citenamefont{Levitov et~al.}(1996)\citenamefont{Levitov, Lee, and
  Lesovik}}]{LevitovLeeLesovik96}
\bibinfo{author}{\bibfnamefont{L.~S.} \bibnamefont{Levitov}},
  \bibinfo{author}{\bibfnamefont{H.~W.} \bibnamefont{Lee}}, \bibnamefont{and}
  \bibinfo{author}{\bibfnamefont{G.}~\bibnamefont{Lesovik}},
  \bibinfo{journal}{J. Math. Phys.} \textbf{\bibinfo{volume}{37}},
  \bibinfo{pages}{4845} (\bibinfo{year}{1996}).

\bibitem[{\citenamefont{Ivanov et~al.}(1997)\citenamefont{Ivanov, Lee, and
  Levitov}}]{IvaLL97}
\bibinfo{author}{\bibfnamefont{D.~A.} \bibnamefont{Ivanov}},
  \bibinfo{author}{\bibfnamefont{H.~W.} \bibnamefont{Lee}}, \bibnamefont{and}
  \bibinfo{author}{\bibfnamefont{L.~S.} \bibnamefont{Levitov}},
  \bibinfo{journal}{Phys. Rev. B} \textbf{\bibinfo{volume}{56}},
  \bibinfo{pages}{6839} (\bibinfo{year}{1997}).

\bibitem[{\citenamefont{d'Ambrumenil and Muzykantskii}(2005)}]{dAMuz05}
\bibinfo{author}{\bibfnamefont{N.}~\bibnamefont{d'Ambrumenil}}
  \bibnamefont{and}
  \bibinfo{author}{\bibfnamefont{B.}~\bibnamefont{Muzykantskii}},
  \bibinfo{journal}{Phys. Rev. B} \textbf{\bibinfo{volume}{71}},
  \bibinfo{pages}{045326} (\bibinfo{year}{2005}).

\bibitem[{\citenamefont{Vanevic et~al.}(2007)\citenamefont{Vanevic, Nazarov,
  and Belzig}}]{VanevicNazBelz07}
\bibinfo{author}{\bibfnamefont{M.}~\bibnamefont{Vanevic}},
  \bibinfo{author}{\bibfnamefont{Y.~V.} \bibnamefont{Nazarov}},
  \bibnamefont{and} \bibinfo{author}{\bibfnamefont{W.}~\bibnamefont{Belzig}},
  \bibinfo{journal}{Phys. Rev. Lett.} \textbf{\bibinfo{volume}{99}},
  \bibinfo{pages}{076601} (\bibinfo{year}{2007}).

\bibitem[{\citenamefont{Keeling et~al.}(2006)\citenamefont{Keeling, Klich, and
  Levitov}}]{KeelingKlichLev06}
\bibinfo{author}{\bibfnamefont{J.}~\bibnamefont{Keeling}},
  \bibinfo{author}{\bibfnamefont{I.}~\bibnamefont{Klich}}, \bibnamefont{and}
  \bibinfo{author}{\bibfnamefont{L.~S.} \bibnamefont{Levitov}},
  \bibinfo{journal}{Phys. Rev. Lett.} \textbf{\bibinfo{volume}{97}},
  \bibinfo{pages}{116403} (\bibinfo{year}{2006}).

\bibitem[{not()}]{note_on_integer_Nu}
\bibinfo{note}{Non-integer $N_u$ would imply a fractional total number of
  occupied states invalidating the mirror-plane property ($\widetilde{h}^2=1$)
  at $T=0$. In practice, experiments would work with zero mean voltage over
  some finite measuring time $t_0$ thereby ensuring $N_u=0$.}

\bibitem[{\citenamefont{Pratap}()}]{Pratap08}
\bibinfo{author}{\bibfnamefont{A.}~\bibnamefont{Pratap}}, \bibinfo{note}{thesis
  (unpublished), University of Warwick, (2008)}.

\bibitem[{\citenamefont{Nozieres and De~Dominicis}(1969)}]{ND69}
\bibinfo{author}{\bibfnamefont{P.}~\bibnamefont{Nozieres}} \bibnamefont{and}
  \bibinfo{author}{\bibfnamefont{C.~T.} \bibnamefont{De~Dominicis}},
  \bibinfo{journal}{Physical Review} \textbf{\bibinfo{volume}{178}},
  \bibinfo{pages}{1079} (\bibinfo{year}{1969}).

\bibitem[{\citenamefont{Anderson}(1967)}]{Anderson67}
\bibinfo{author}{\bibfnamefont{P.}~\bibnamefont{Anderson}},
  \bibinfo{journal}{Phys. Rev. Lett.} \textbf{\bibinfo{volume}{18}}
  (\bibinfo{year}{1967}).

\bibitem[{\citenamefont{Levitov and Lesovik}(1993)}]{LevitovJETP93}
\bibinfo{author}{\bibfnamefont{L.}~\bibnamefont{Levitov}} \bibnamefont{and}
  \bibinfo{author}{\bibfnamefont{G.}~\bibnamefont{Lesovik}},
  \bibinfo{journal}{JETP Letters} \textbf{\bibinfo{volume}{58}},
  \bibinfo{pages}{230} (\bibinfo{year}{1993}).

\bibitem[{dAM()}]{dAMuz08}
\bibinfo{note}{The relation of the FCS to the geometry of the two planes
  corresponding to $\widetilde{h}$ and $h$ is more complicated if the
  transformation mixes left and right states. However, in the equilibrium case
  it is possible to switch to a basis of scattering states which diagonalizes
  the scattering matrix $S$ and identify the corresponding angles easily.}

\end{thebibliography}

\end{document}